\documentclass[11pt,a4paper]{article}
\pdfoutput=1 
\usepackage{jheppub}	
\usepackage{tikz}
\usetikzlibrary{patterns}
\usetikzlibrary{positioning,arrows}
\usetikzlibrary{decorations.pathmorphing}
\usetikzlibrary{decorations.markings}
\usetikzlibrary{snakes}
\usetikzlibrary{matrix}

\usepackage{amsmath}
\usepackage{bm}
\usepackage[utf8]{inputenc}
\usepackage[english]{babel}
\usepackage{makeidx}
\usepackage{tikz}
\tikzset{every picture/.style={line width=0.75pt}} 
\usepackage{amsfonts}
\usepackage{enumerate}
\usepackage{mathrsfs}
\usepackage{tensor}
\usepackage[autostyle]{csquotes}
\usepackage{subfig}

\def\be{\begin{equation}}
	\def\ee{\end{equation}}
\def\bea{\begin{eqnarray}}
	\def\eea{\end{eqnarray}}
\newcommand{\nn}{\nonumber}

\def\apjl{\ref@jnl{ApJ}}

\newcommand{\RNum}[1]{\uppercase\expandafter{\romannumeral #1\relax}}

\usepackage{amsmath,amssymb}
\usepackage{latexsym}
\usepackage{graphicx}
\usepackage{slashed}

\usepackage[vcentermath,enableskew]{youngtab}
\usepackage{epsfig}

\def\be{\begin{equation}}
	\def\ee{\end{equation}}
\def\bea{\begin{eqnarray}}
	\def\eea{\end{eqnarray}}

\title{On 5-point conformal block with level 2 degenerate field insertion and its 
	AGT dual}

\author[]{Hasmik Poghosyan and}
\author[]{Rubik Poghossian}
\emailAdd{h.poghosyan@yerphi.am}
\emailAdd{poghos@yerphi.am}
\affiliation[]{Yerevan Physics Institute \\
	Alikhanian Br. 2, 0036 Yerevan, Armenia}

\abstract{In this paper, we develop and explore recursive methods to investigate the 2d CFT 5-point conformal block with a level 2 degenerate insertion, as well as its AGT dual, by solving the BPZ differential equation.
	First, we represent the solution of the differential equation as a double series expansion. On the 2-node quiver gauge theory side, this corresponds to the instanton series. We then demonstrate that the  expansion coefficients are uniquely determined by a recursion relation.
	Inspired by the 
	approach initiated in a paper by D. Gaiotto and J. Teschner, we partially resum this series and show that 
	the result can be elegantly expressed in terms of a single hypergeometric function and 
	its derivative. This new representation makes it straightforward to relate different 
	asymptotic regions. As a by-product, this provides us a simple derivation of fusion and 
	braiding coefficients. 
	
	We describe the subtle procedure of merging the degenerate field with the 
	outgoing state, thereby obtaining a generic 4-point block, which on the gauge theory 
	side  corresponds to the partition function of $SU(2)$ gauge theory with four 
	massive hypermultiplets in the $\Omega$-background. 
	
	Finally, we performed several nontrivial checks,  which confirm our results.
	
}
 \clearpage
\begin{document}
	\tikzset{
		line/.style={thick, decorate, draw=black,}
	}
	
	\maketitle
	
	
	\section{Introduction}
	Initially, a gravitational background, referred to as the $\Omega$ background, was introduced to localize quantum field theory  path integrals of supersymmetry-protected observables around a discrete set of configurations, thereby, making these quantities explicitly calculable.
	
	One recovers the initial Seiberg-Witten  theory  in flat spacetime \cite{Seiberg:1994rs,Seiberg:1994aj} by simply sending the $\Omega$-background parameters, traditionally denoted by $\epsilon_1$ and $\epsilon_2$, to zero. The development of localization techniques \cite{Nekrasov:2002qd,Flume:2002az,Bruzzo:2002xf,Nekrasov:2003rj,Flume:2004rp} has provided  easy access  to explicit computations.
	
	That the significance of  $\Omega$-background is much bigger than a 
	convenient mathematical regularization tool became clear after 
	discovery of the striking fact that 2d CFT correlation functions are closely
	related to gauge theory partition function in $\Omega$-background (AGT duality) 
	\cite{Gaiotto:2009we,Alday:2009aq,Poghossian:2009mk}.

	In \cite{Gaiotto:2012sf}, a useful recursive method for calculating Liouville CFT blocks was proposed. 
	This approach involves examining the 5-point conformal block with a level two degenerate field insertion,
	which satisfies a second-order differential equation, as first introduced by Belavin, Polyakov, and Zamolodchikov \cite{Belavin:1984vu}.
	The solution of this differential equation can be expressed as a double series, where the coefficients satisfy a recursion relation. 
	It has been argued that this double sum can  be represented in terms of differential operators acting on a hypergeometric function. The main result of this paper is the development of this idea by showing that the differential operator under consideration is, in fact, a first-order operator. We derive an efficient 
	recursion relation, which allows one to compute the coefficients of this differential operator.

	Since the conformal block is expressed in terms of hypergeometric function, the analytic 
	continuation along contours that connect different singularities becomes an easy task. In particular, 
	braiding and fusion coefficients can be readily deduced from hypergeometric function identities. 
	Furthermore, taking the limits when a degenerate field collides with any of the 
	other primary fields, provides an alternative way to compute  4-point block or its 
	AGT dual, namely the Nekrasov partition function of $SU(2)$ gauge theory with four hypers.    
	
	The paper is organized as follows:  
	
	In section \ref{5pntBlock}, we write the BPZ differential equation
	satisfied by the five
	point conformal block with a level two degenerate field insertion.
	This five point function can be written as a double series in the two insertion
	points and we find a recursion relation for the coefficient of this series. 
	Upon examining this recursion relation, we notice that there 
	are two different ways to represent the sum
	related with the insertion point of the degenerate field.
	Both represent first order differential
	operators acting on distinct  $_2F_1$    hypergeometric functions.
	
	In section \ref{br_fus}, we demonstrate that
	these analytic representations of the insertion point of the degenerate
	field allow  one to obtain the braiding and fusion matrices, as well as  the four point conformal block. 
	Depending whether one recovers the four point conformal
	block by merging the degenerate field, with the state at infinity
	or zero, one has to choose  the respective analytic expression.
	We demonstrate this procedure in full details.  
	
	Appendix \ref{instAGT} is a  review on  instanton counting 
	and the AGT duality.
	
	\section{Liouville degenerate $5$-point conformal block}
	\label{5pntBlock}
	Our conventions on 2d Liouville theory are standard. The Virasoro central charge 
	is parametrized as 
	\be
	c=1+6Q^2\,\,,\qquad Q=b+b^{-1}
	\ee
	with $b$ being the dimensionless coupling 
	of Liouville theory. The conformal dimensions of primary fields/states are related to 
	the momentum parameters as
	\bea
	h_p=\frac{Q^2}{4}-p^2
	\eea
	The five-point function with a second level degenerate insertion
	\bea
	\label{5pointCB}
	{\cal F}(z,x)=\langle p_0| V_{k_0} (1) V_{k_{deg}}(z) V_{k_2} (x)  |p_3\rangle
	\eea
	satisfies the BPZ \cite{Belavin:1984vu} differential equation
	\begin{small}
		\bea
		\label{BPZ}
		& \left(
		\frac{1}{b^2}\frac{\partial ^2}{\partial z^2}-\frac{2 z-1}{z (z-1)} \frac{\partial }{\partial z}+
		\frac{x (x-1) }{z (z-1) (z-x)}\frac{\partial }{\partial x}+
		\left(\frac{h_{k_2}}{(z-x)^2}+\frac{h_{p_3}}{z^2}+\frac{h_{k_0}}{(z-1)^2}-\frac{\delta }{z (z-1)}\right)
		\right){\cal F}(z,x)=0
		\eea
	\end{small}
	where 
	\bea
	\delta =-h_{p_0}+h_{k_0}+h_{k_{deg}}+h_{k_2}+h_{p_3}
	\eea
	and 
	$V_{k_{deg}}(z)$  is the level two 
	degenerate field with momentum parameter 
	\bea
	\label{deg_field}
	k_{deg}= b+\frac{1}{2 b}
	\eea
	OPE fusion rules impose restrictions on the intermediate field dimensions:
	\bea
	\label{OPEint}
	p_1= p-\frac{b \sigma }{2}\,,
	\quad
	p_2= p
	\quad
	{\rm where }
	\quad
	\sigma=\pm 1
	\eea	
	\subsection{Modified BPZ differential equation and its solution}
	\label{r0}
	We   separate the free field part from the conformal block by  introducing 
	\bea
	\label{GvsF}
	G(z,x)=A^{-1}(z,x)	{\cal F}(z,x) 
	\eea
	where
	\begin{footnotesize}
		\bea
		\label{Azx}
		A(z,x)=
		(1-x)^{-\frac{1}{2} \left(Q-2 k_0\right) \left(Q-2 k_2\right)}x^{-\frac{1}{2} \left(Q-2 k_2\right) \left(Q-2 p_3\right)} (1-z)^{\frac{b}{2}  \left(Q-2 k_0\right)} z^{\frac{b}{2}  \left(Q-2 p_3\right)}  (z-x)^{\frac{b}{2}  \left(Q-2 k_2\right)}
		\eea
	\end{footnotesize}
	The newly defined function $G(x,z)$ satisfies the differential equation
	\begin{small}
		\bea
		\label{BPZren}
		& \left(
		\frac{\partial ^2}{\partial z^2}
		+\left(\frac{\kappa _2+\kappa _3}{z-1}+\frac{1-\gamma _0+2 \kappa _1}{z-x}+\frac{\gamma _0}{z}\right)\frac{\partial }{\partial z}
		+\frac{b^2 x (x-1)}{z (z-1) (z-x)}\frac{\partial }{\partial x}
		+ \frac{\left(\kappa _1+\kappa _2\right) \left(\kappa _1+\kappa _3\right)}{z (z-1) }
		\right)G(z,x)=0\qquad
		\eea
	\end{small}
	with
	\bea
	\label{kappaToP}
	\gamma _0 &=& 1-2 b p_3 \,,
	\qquad
	\kappa _1 = \frac{b}{2} \left(Q-2  k_2-2  p_3\right)\,,
	\\
	\kappa _2 &=&  \frac{1}{2} \left(1-2 b k_0-2 b p_0\right)\,,
	\qquad
	\kappa _3 = \frac{1}{2} \left(1-2 b k_0+2 b p_0\right)
	\eea
	Either from the OPE structure or from AGT and instanton counting, 
	it can be observed that the differential equation (\ref{BPZren}) admits a solution in the form of a double series
	\bea
	\label{solreg}
	G(z,x)=z^{r}x^{s}\sum_{j\ge 0,j+i\ge 0}^{\infty}z^{i} x^j d_{i,j}
	\eea
	valid in the region
	\bea
	|z|< 1,
	\quad
	|x/z|< 1
	\eea
	By inserting (\ref{solreg}) in (\ref{BPZren}) one derives  the following relation 
	among the expansion coefficients
	\bea
	\label{receq0}
	A_0(i,j) d_{i,j}+A_1(i) d_{i-1,j}
	+A_2(i,j) d_{i,j-1}
	+A_3(i) d_{i+1,j-1}=0
	\eea
	where
	\bea
	A_0(i,j)=-b^2 (j+s)-2 \kappa _1 (i+r)-(i+r)^2
	\\
	A_1(i)=\left(i+\kappa _1+\kappa _2+r-1\right) \left(i+\kappa _1+\kappa _3+r-1\right)
	\\ \nn
	A_2(i,j)=b^2 (j+s-1)-\gamma _0 (i+r)-\left(\kappa _2+\kappa _3\right) (i+r)
	\\ 
	-(i+r-1) (i+r) -\left(\kappa _1+\kappa _2\right) \left(\kappa _1+\kappa _3\right)
	\\
	A_3(i)=(i+r+1) \left(\gamma _0+i+r\right)
	\eea
	We normalize the solution by setting $d_{0,0} = 1$. Then consistency with the conditions
	\bea
	\label{dij_initial}
	d_{0,0}=1\,;
	\quad
	d_{i,j}=0
	\quad
	{\rm if}
	\quad
	j< 0
	\quad
	{\rm  or}\quad
	j+i< 0
	\eea
	requires that $A_0(0,0)=0$ hence
	\bea
	s=-\frac{r \left(2 \kappa _1+r\right)}{b^2}
	\eea
	\subsection{Recursive solution}
	\begin{figure}[t]
		\centering
		\begin{tikzpicture}[scale=0.50]
			\draw[step=1cm,gray,dotted,very thin] (0,-0.5) grid (10,8);
			\draw[->] (1,4)--(8,4) node[right]{$j$};
			\path[pattern={north east lines},pattern color= lightgray]  (3,4) --  (3,1)-- (6,1)-- (6,7);
			\draw[ color=black] (3,4)-- (6.5,7.5);
			\draw[<-] (3,0.3)--(3,7.5) ;
			\node at (3,-0.1){i};
			\draw[ thick,color=gray]  (2,3)--(2,4)--(3,5)--(3,4)--(2,3);
			\foreach \p in {(2,4),(2,3), (3,5) }
			\fill[orange] \p circle(.08);
			\fill[blue]  (3,4)  circle(.16);
			\fill[orange]  (3,4)  circle(.12);
			\foreach \p in {(3,1),(3,2),(3,3),(4,1),(4,2),(4,3),(4,4),(4,5),
				(5,1),(5,2),(5,3),(5,4),(5,5),(5,6),  (6,1),(6,2),(6,3),(6,4),(6,5),(6,6),(6,7)}
			\fill[blue] \p circle(.12);
		\end{tikzpicture}
		\caption{The  part that is not shadowed corresponds to the values of $i$ and $j$ for which $d_{i,j}=0$. On the boundaries we have (\ref{dmjj_0}) and (\ref{di0_0}).
			Using   (\ref{recrel0}) the values in the shaded region can be successively determined by   transporting   the parallelogram with orange vertices.
		}
		\label{fig:r0}
	\end{figure}
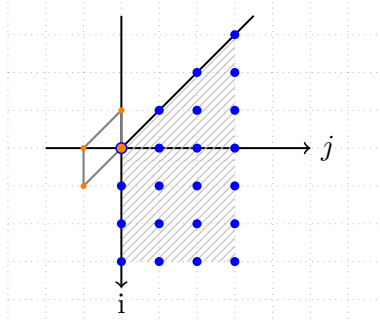
	Let us rewrite the equation (\ref{receq0}) in form of a double index recursion relation: 
	\bea
	\label{recrel0}
	d_{i,j}=-\frac{A_1(i)}{A_0(i,j)}d_{i-1,j}
	-\frac{A_2(i,j)}{A_0(i,j)}d_{i,j-1} 
	-\frac{A_3(i)}{A_0(i,j)}d_{i+1,j-1}
	\eea
	The Fig.\ref{fig:r0} makes it apparent that successive application of this relation
	together with (\ref{dij_initial}) uniquely defines all coefficients $d_{i,j}$
	belonging to the shaded region.

	To be consistent with  the $z\to 0$ OPE fusion rules, the parameter $r$ should take one of two possible values:  
	\bea
	r=
	b p \sigma -\kappa _1
	\eea
	where $\sigma=\pm 1$ and $p=p_2$ is the momentum parameter of the intermediate state (see (\ref{OPEint})). 
	For the boundary points of the shaded region, as shown in Fig.\ref{fig:r0}, the recursion relation described in equation (\ref{recrel0}) leads to particularly simple expressions
	\bea
	\label{dmjj_0}
	d_{-j,j}=
	\frac{(-r)_j \left(1-r-\gamma _0\right)_j}{j! \left(b Q-2 r-2 \kappa _1\right)_j}
	\eea
	and
	\bea
	\label{di0_0}
	d_{i,0}=
	\frac{\left(r+\kappa _1+\kappa _2\right)_i \left(r+\kappa _1+\kappa _3\right)_i}{i! \left(2 r+2 \kappa _1+1\right)_i}
	\eea
	which play an important role in what follows. According to the same recursion relation we have
	\bea
	&b^2 d_{0,1}=
	\frac{(r+1) \left(\gamma _0+r\right) \left(\kappa _1+\kappa _2+r\right) \left(\kappa _1+\kappa _3+r\right)}{2 \kappa _1+2 r+1}
	-\frac{r \left(\gamma _0+r-1\right) \left(\kappa _1+\kappa _2+r-1\right) \left(\kappa _1+\kappa _3+r-1\right)}{2 \kappa _1+2 r- b Q}
	\\ \nn
	&\qquad \quad -\left(\kappa _1+\kappa _2\right) \left(\kappa _1+\kappa _3\right)
	-\gamma _0 r-r \left(2 \kappa _1+r\right)-\left(\kappa _2+\kappa _3\right) r+(1-r) r
	\eea
	We have checked that when these values are inserted into (\ref{solreg}), 
	one gets consistent with instanton counting result (see Appendix \ref{instAGT}). 
	\subsubsection{Solution in terms of hypergeometric function}
	Observe that the sum (\ref{solreg}) can be rearranged in two alternative ways
	\bea
	\label{G1firstApr_r0}
	G(z,x)
	&=&z^{r}x^{s}\sum_{j=0}^{\infty}x^j\sum_{l=0}^\infty d_{l-j,j}z^{l-j}\\
	\label{G2firstApr_r0}
	&=&y^{-r}x^{r+s}\sum_{j=0}^{\infty}x^j\sum_{l=0}^\infty d_{j-l,l}y^{l-j}
	\eea
	where we have introduced notation 
	\[
	y=\frac{x}{z}
	\]
	From now on we will adopt a more detailed notation $G(z,x|p_1,p_2)$ for the function $G(z,x)$, explicitly 
	indicating intermediate state momenta $p_1$ and $p_2$. 
	From (\ref{dmjj_0}) and  (\ref{di0_0}), for the most important at $x\to 0$ limit contributions in (\ref{G1firstApr_r0}) and (\ref{G2firstApr_r0}), we get
	\bea 
	&&G(z,x|p^{-\sigma},p)|_{x\to 0,z\,\text{fixed}}\sim x^{\frac{\kappa _1^2}{b^2}-p^2}H^{1}_\sigma (z)\\
	&&G(z,x|p,p^\sigma)|_{x\to 0,y\,\text{fixed}}\sim x^{\left(\frac{\kappa _1}{b}-\frac{b}{2}\right){}^2-p^2}H^{2}_\sigma (y)
	\eea 
	where
	\bea
	\label{H1}
	&&H^{1}_\sigma(z)
	=
	z^{b p \sigma -\kappa _1} \, _2F_1\left(b p \sigma +\kappa _2,b p \sigma +\kappa _3;2 b p \sigma +1;z\right)\\
	\label{H2}
	&&H^{2}_\sigma(y)
	=y^{ \kappa _1-b p \sigma-\frac{b^2}{2}} \, _2F_1\left(\kappa _1-p \sigma  b-\frac{b^2}{2},\kappa _1-p \sigma  b-\gamma _0-\frac{b^2}{2}+1;1-2 b p \sigma ;y\right)\qquad
	\eea
	We have also used the notation
	\bea
	p^\sigma= p+ \frac{\sigma b}{2}
	\eea
	Exploring recursion (\ref{recrel0}) we have noticed that for fixed $j$, 
	the coefficients $d_{l,j}$ ($d_{j-l,l}$) can be expressed in terms of  $d_{l,0}$ ($d_{-l,l}$).
	This observation, together with (\ref{H1}) and (\ref{H2}), led us to the ans{\"a}tze
	\bea
	\label{Ganz1}
	&&G(z,x|p^{-\sigma},p)=x^{\frac{\kappa _1^2}{b^2}-p^2}(P_1(x,z)H^1_\sigma(z)
	+\hat{P_1}(x,z)z H^{1 \, '}  _{\sigma}   (z))
	\\
	\label{Ganz2}
	&&G(z,x|p,p^\sigma)=x^{\left(\frac{\kappa _1}{b}-\frac{b}{2}\right){}^2-p^2}(P_2(x,y)H^2_\sigma(y)+\hat{P_2}(x,y) y H^{2 \, '}_\sigma(y))
	\eea
	where 
	\bea
	\label{cij}
	P_1(x,z)=\sum_{i=0}^\infty \sum_{j=0}^{i}h^{(1)}_{i,j}x^i z^{j-i}\,,
	\quad
	\hat{P_1}(x,z)=\sum_{i=0}^\infty \sum_{j=0}^{i}\hat{h}^{(1)}_{i,j}x^i z^{j-i}\\
	\label{hij}
	P_2(x,y)=\sum_{i=0}^\infty \sum_{j=0}^{i}h^{(2)}_{i,j}x^i y^{j-i}\,,
	\quad
	\hat{P_2}(x,y)=\sum_{i=0}^\infty \sum_{j=0}^{i}\hat{h}^{(2)}_{i,j}x^i y^{j-i}
	\eea 
	By imposing differential equation (\ref{BPZren}) on (\ref{Ganz1}) and 
	using the identity
	\bea 
	&H_{\sigma }^{1''}(z)=-\frac{b^2 p^2-\kappa _1^2+\left(\kappa _1+\kappa _2\right) \left(\kappa _1+\kappa _3\right) z}{(z-1) z^2}H_{\sigma }^1(z) +\frac{1+2 \kappa _1-\left(2 \kappa _1+\kappa _2+\kappa _3+1\right) z}{(z-1) z} H_{\sigma }^{1'}(z)
	\eea 
	to reduce higher order derivatives to $H_\sigma (z)$ and $H_{\sigma }^{1'}(z)$, 
	we have obtained recursion relations. They take a much simpler form if one introduces the 
	linear combinations
	\bea
	h^{(1)}_{i,j}=\sum_{\sigma=\pm 1}\left(\kappa _1+ \sigma b  p\right) t_{i,j}(\sigma p)\,,
	\quad
	\hat{h}^{(1)}_{i,j} = \sum_{\sigma=\pm 1} t_{i,j}(\sigma p)
	\eea
	Then the recursion relations for coefficients $t_{i,j}(\sigma p)$ have the form
	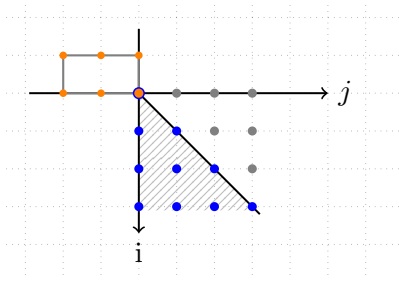
\begin{figure}[t!]
		\centering
		\begin{tikzpicture}[scale=0.50]
			\draw[step=1cm,gray,dotted,very thin] (-0.5,-0.8) grid (10,6.6);
			\path[pattern={north east lines},pattern color= lightgray]  (3,4) --  (3,0.9)-- (6,0.9)-- (6,1);
			\draw (6.2,0.8)--(3,4);
			\draw[->] (0.1,4)--(8,4) node[right]{$j$};
			\draw[<-] (3,0.3)--(3,5.7) ;
			\node at (3,-0.2){i};
			\draw[thick, color=gray]  (1,4)--(3,4)--(3,5)--(1,5)--(1,4);
			\foreach \p in {(1,4),(2,4),(1,5),(2,5), (3,5) }
			\fill[orange] \p circle(.1);
			\fill[blue]  (3,4)  circle(.16);
			\fill[orange]  (3,4)  circle(.12);
			\foreach \p in {(3,1),(3,2),(3,3),(4,1),(4,2),(4,3),(5,1),(5,2),(6,1)}
			\fill[blue] \p circle(.12);
			\foreach \p in {(6,2),(5,3),(6,3),(4,4),(5,4),(6,4)}
			\fill[gray] \p circle(.12);
		\end{tikzpicture}
		\caption{The  part that is not shadowed corresponds to the values of $i$ and $j$ where $t_{i,j}=0$. The values on the  vertical  boundary are given in (\ref{ti0}).}
		\label{fig:tRec}
	\end{figure}
	\bea
	\label{RecrelTr0}
	B_{(0,0)(i,j)}(p)t_{i,j}(p)=
	\sum_{\substack{s \in \{0,1\},l\in{\{0,1,2\}} \\ (s,l)\neq (0,0)}} 
	B_{(s,l),(i,j)}(p)t_{i-s,j-l}(p)
	\\ \nn
	+\hat{B}_{(s,l),(i,j)}(p)t_{i-s,j-l}(-p)
	\eea
	where
	\bea
	\label{B_Bhat}
	B_{(0,0)(i,j)}(p)&=&4 b^2 p^2 (j-i)+2 b p \left(b^2 i+(i-j)^2\right)
	\\
	B_{(1,0)(i,j)}(p)&=&2 b p \left(b p-i+j-\kappa _1+1\right) \left(b p+\gamma _0-i+j-\kappa _1\right)
	\eea
	\bea
	B_{(0,1)(i,j)}(p)=b p \left(2 i \left(b^2-4 j+4\right)+4 i^2+4 (j-1)^2-\kappa _2-\kappa _3\right)
	\nn\\
	-b^2 p^2 (6 (i-j)+7)+\kappa _2 \kappa _3 (2 (i-j)+1)
	\eea
	\bea
	\label{B02}
	B_{(0,2)(i,j)}(p)=2 (i-j+2) \left(b p (b p-i+j-2)-\kappa _2 \kappa _3\right)
	\eea
	\begin{small}
		\bea
		B_{(1,1)(i,j)}(p)=-4 b^3 p^3	+b^2 p^2 \left(-3 \gamma _0+6 (i-j)+6 \kappa _1+4\right)+ b \gamma _0 p \left(4 (i-j)+4 \kappa _1+\kappa _2+\kappa _3\right)
		\nn\\
		+b p \left(2 b^2 (i-1)-2 \kappa _1 \left(4 (i-j)+2 \kappa _1+\kappa _2+\kappa _3+2\right)-4 (i-j) (i-j+1)\right)
		\nn\\
		+\kappa _2 \kappa _3 \left(\gamma _0-2 \left(i-j+\kappa _1\right)\right)
		\quad
		\eea
		\bea
		B_{(1,2)(i,j)}(p)=	2 b^3 p^3+b^2 p^2 \left(\gamma _0-2 i+2 j-2 \kappa _1-3\right)-b \gamma _0 p \left(2 i-2 j+2 \kappa _1+\kappa _2+\kappa _3+2\right)
		\nn\\
		+b p \left(2 \kappa _1 \left(2( i- j)+\kappa _1+\kappa _2+\kappa _3+3\right)-2 b^2 (i-1)+2 (i-j+1) (i-j+2)+\kappa _2+\kappa _3\right)
		\nn\\
		+\kappa _2 \kappa _3 \left(-\gamma _0+2 i-2 j+2 \kappa _1+3\right)\quad
		\eea
	\end{small}
	\bea
	\hat{B}_{(1,0)(i,j)}(p)&=&0\,,
	\\
	\hat{B}_{(0,1)(i,j)}(p)&=&(2 (i-j)+1) \left(b p+\kappa _2\right) \left(b p+\kappa _3\right)
	\\
	\label{Bh02}
	\hat{B}_{(0,2)(i,j)}(p)&=&-2 (i-j+2) \left(b p+\kappa _2\right) \left(b p+\kappa _3\right)
	\\
	\hat{B}_{(1,1)(i,j)}(p)&=&\left(b p+\kappa _2\right) \left(b p+\kappa _3\right) \left(\gamma _0-2 \left(i-j+\kappa _1\right)\right)
	\\
	\hat{B}_{(1,2)(i,j)}(p)&=&\left(b p+\kappa _2\right) \left(b p+\kappa _3\right) \left(3-\gamma _0+2( i- j+ \kappa _1)\right)
	\eea
	These relations together with boundary conditions
	\bea
	&&t_{0,0}( p)=\frac{1}{2 b  p} \nn\\
	&&t_{i,j}=0\,,\quad {\rm if} \quad i<0 \quad {\rm  or}\quad
	j<0 
	\nn
	\eea
	uniquely determine the coefficients $t_{i,j}$ 
	(even in the region $0\le i<j$ ). On the other hand, to ensure the proposed 
	ans{\"a}tze (\ref{Ganz1}), all coefficients with $j>i$ must vanish. Besides, this 
	condition is crucial to get a correct OPE behaviour in $z\to \infty $ limit (see subsection \ref{4pointBlock}).
	Evidently setting the coefficients $0$ for $j>i+2$ is trivially consistent with (\ref{RecrelTr0}). 
	Also the case $j=i+2$  does not cause any problem, since as it is evident from equations (\ref{B02}) and (\ref{Bh02})
	\bea
	B_{(0,2)(i,i+2)}=\hat{B}_{(0,2),(i,i+2)}=0
	\eea
	Thus the only problematic case to be checked is 
	the case $j=i+1$. Vanishing of coefficients on this diagonal, from (\ref{RecrelTr0}), is equivalent to conditions 
	that $t_{i,i+1}=0$ which according to (\ref{RecrelTr0}) is equivalent to
	\bea
	\label{extra_conditions}
	0&=&B_{(0,1),(i,i+1)}(p)t_{i,i}(p)+B_{(0,2),(i,i+1)}(p)t_{i,i-1}(p)+B_{(1,2),(i,i+1)}(p)t_{i-1,i-1}(p)
	\\ \nn
	&+&\hat{B}_{(0,1),(i,i+1)}(p)t_{i,i}(-p)+\hat{B}_{(0,2),(i,i+1)}(p)t_{i,i-1}(-p)+\hat{B}_{(1,2),(i,i+1)}(p)t_{i-1,i-1}(-p)
	\eea	
	Though  we do not have a general proof, automated calculations with the help of Wolfram Mathematica,  
	up to order  $i=100$, confirms that this condition is indeed satisfied.	
	
	Using the recursion relation we have noticed that
	\bea
	\label{ti0}
	t_{i,0}(p)=\frac{\left(\kappa _1-b p\right)_i \left(1-b p-\gamma _0+\kappa _1\right)_i}{2 b p\left(b Q-2 p b\right)_i i! }
	\eea
	Here is the first nontrivial coefficient not given by (\ref{ti0})
	\begin{small}
		\bea
		\nn
		4 b^4 p^2 t_{1,1}=
		(b Q-2 b p)^{-1}\left(b p-\kappa _1\right) \left(b p \left(2 b^2-7 b p-\kappa _3+4\right)+\kappa _2 \left(\kappa _3-b p\right)\right) \left(b p+\gamma _0-\kappa _1-1\right)
		\\ \nn
		+b p \left(\gamma _0 \left(4 \kappa _1+\kappa _2-3 b p\right)-2 \kappa _1 \left(2 \kappa _1+\kappa _2-3 b p+2\right)-4 b p (b p-1)\right)
		\\ \nn
		-(b Q+2 b p)^{-1}\left(b p+\kappa _1\right) \left(b p+\kappa _2\right) \left(b p+\kappa _3\right) \left(b p-\gamma _0+\kappa _1+1\right)
		\\
		+\kappa _3 \left(\gamma _0-2 \kappa _1\right) \left(b p+\kappa _2\right)-\left(\gamma _0-2 \kappa _1\right) \left(b p+\kappa _2\right) \left(b p+\kappa _3\right)
		\qquad
		\eea
	\end{small}
	As will be shown in the next subsection
	$h^{(2)}_{i,j}$ and 
	$\hat{h}^{(2)}_{i,j}$  given in (\ref{hij})  can be found from $h^{(1)}_{i,j}$ and $\hat{h}^{(1)}_{i,j}$ through a simple redefinition of parameters 
	corresponding to conformal transformation $z\to x/z$. 
	\subsection{Symmetries}
	Using recursion relation (\ref{recrel0}) it is easy to check that there is a symmetry
	\bea
	d_{i-j,j}   \to   d_{j-i,i}
	\eea
	provided we simultaneously replace
	\bea
	&&k_0\to k_2\,,
	\quad
	k_2\to k_0
	\\
	&&p_3\to p_0\,,
	\quad
	p_0\to p_3
	\\
	&&p\to p- \frac{\sigma b}{2}\,,
	\quad
	\sigma \to -\sigma
	\eea
	Of course this is not a coincidence and can be easily derived performing 
	conformal transformation $\zeta \to x/\zeta$ which maps insertion points of $5$-point CFT block
	$(\infty,1,z,x,0)$ to $(0,x,x/z,1,\infty)$. The same transformation interchanges the 
	representations (\ref{Ganz1}) and (\ref{Ganz2}). 
	Here is the explicit map:
	\bea
	\label{hcSym_r0}
	h^{(2)}_{i,j} \to h^{(1)}_{i,j}-\left(\kappa _1+\kappa _2\right) \hat{h}^{(1)}_{i,j}\,,
	\quad
	\hat{h}^{(2)}_{i,j} \to \hat{h}^{(1)}_{i,j}
	\eea
	provided one redefines the parameters as follows:
	\bea
	\gamma _0 &\to& \kappa _2-\kappa _3+1,
	\quad
	\kappa _1\to \kappa _2+\frac{b^2}{2}
	\\
	\kappa _2 &\to&\kappa _1 -\gamma _0+1 -\frac{b^2}{2},
	\quad
	\kappa _3\to \kappa _1-\frac{b^2}{2}
	\eea
	\section{Braiding and Fusion matrices}
	\label{br_fus}
	Given the representations (\ref{Ganz1}) and (\ref{Ganz2}), the analytic continuation of the conformal block
	from one region to another reduces to that of hypergeometric functions, which 	effectively carry all the singularity structure.
	With the help of following standard hypergeometric identities
	\bea 
	\label{rule2F11}
	\, _2F_1(a,b;c;z)&=&\frac{\Gamma (c) \Gamma (c-a-b) 
	}{\Gamma (c-a) \Gamma (c-b)}
	\,_2F_1(a,b;a+b-c+1;1-z)
	\\ \nn
	&+&\frac{(1-z)^{c-a-b} \Gamma (c) \Gamma (a+b-c)}{\Gamma (a) \Gamma (b)}\,
	_2F_1(c-a,c-b;c-a-b+1;1-z)
	\eea 
	and
	\bea
	\label{rule2F1inf}
	\, _2F_1(a,b;c;z) &=&
	\frac{ \Gamma (c) \Gamma (b-a)}{ \Gamma (b) \Gamma (c-a)}(-z)^{-a} 
	\, _2F_1(a,a-c+1;a-b+1;z^{-1})
	\\
	\nn
	&+&
	\frac{\Gamma (c) \Gamma (a-b)}{ \Gamma (a) \Gamma (c-b)}(-z)^{-b}
	\, _2F_1(b,1 + b - c;1 + b - a;z^{-1})
	\eea
	instead of  (\ref{Ganz1}) and (\ref{Ganz2}), we can find two other pairs of solutions of the differential equation (\ref{BPZren}). Namely, we apply the identity (\ref{rule2F11}) to (\ref{Ganz1})  ((\ref{rule2F1inf}) to (\ref{Ganz2})) and separate the parts with definite monodromies around $z\sim 1$ ($y\sim \infty $) and get the solutions
	\bea
	\label{Ganz3}
	&&G_{{\text f}}(z,x|k_0^{\sigma},p)=x^{\frac{\kappa _1^2}{b^2}-p^2}(P_1(x,z){\tilde H}^1_\sigma(z)
	+\hat{P_1}(x,z)z {\tilde H}^{1 \, '}  _{\sigma}   (z))\,,
	\\
	\label{Ganz4}
	&&G_{{\text b}}(z,x|p,p_3^{-\sigma})=x^{\left(\frac{\kappa _1}{b}-\frac{b}{2}\right){}^2-p^2}(P_2(x,y){\tilde H}^2_\sigma(y)+\hat{P_2}(x,y) y {\tilde H}^{2 \, '}_\sigma(y))
	\eea
	where
	\begin{small}
		\bea 
		{\tilde H}^1_\sigma(z)&=& z^{b p-\kappa _1}(1-z)^{\frac{1-\sigma}{2} \left(1-\kappa _2-\kappa _3\right)} \,\\ &\times &_2F_1\left(\frac{1}{2}+b p+ \sigma\left(\kappa _2-\frac{1}{2}\right) ,\frac{1}{2}+b p+ \sigma\left(\kappa _3-\frac{1}{2}\right);1+ \sigma\left(\kappa _2+\kappa _3-1\right);1-z\right)\nn\\
		{\tilde H}^2_\sigma(y)&=&y ^{\frac{1+\sigma}{2} \left(\gamma _0-1\right)}
		\left(1-\frac{1}{y }\right)^{\frac{1+\sigma}{2}\left(b^2+\gamma _0-2 \kappa _1\right)}\\
		&\times &\, _2F_1\left(\frac{1}{2}+bp+\sigma  \left(\frac{b Q}{2}-\kappa _1\right),\frac{1}{2}-bp+\sigma  \left(\frac{b Q}{2}-\kappa _1\right);1+\sigma\left(1-\gamma _0\right);\frac{1}{y }\right)\nn
		\eea
	\end{small}
	\begin{figure}[t]
		\begin{center}
			\begin{tikzpicture}[scale=0.50]
				\node[scale=1] at (-0.5,2.2){${\cal F}_{\text f}=$};
				\draw (2,1)    -- (8.5,1);
				\draw (7,1) -- (7,3);
				\draw (3.5,1) -- (3.5,3);
				\draw [dashed] (3.5,2)--(5,3);
				\node[scale=0.8] at (3.5,3.4){$k_0$};
				\node[scale=0.8] at (5.25,3.4){$k_{deg}$};
				\node[scale=0.8] at (7,3.4){$k_2$};
				\node[scale=0.8] at (2.75,0.45){$p_0$};
				\node[scale=0.8] at (5.5,0.45){$p$};
				\node[scale=0.8] at (8,0.45){$p_3$};
				\node[scale=0.8] at (3.9,1.7){$k_0^{\sigma}$};
				\node[scale=0.8] at (1.5,1){$\infty$};
				\node[scale=0.8] at (9,1){$0$};
				\node[scale=1] at (13,2.2){${\cal F}_{\text b}=$};
				\draw (15,1)    -- (21.5,1);
				\draw (16.5,1) -- (16.5,3);
				\draw (18.25,1) -- (18.25,3);
				\draw [dashed] (20,1) -- (20,3);
				\node[scale=0.8] at (16.5,3.4){$k_0$};
				\node[scale=0.8] at (18.25,3.4){$k_2$};
				\node[scale=0.8] at (20,3.4){$k_{deg}$};
				\node[scale=0.8] at (15.75,0.45){$p_0$};
				\node[scale=0.8] at (21,0.45){$p_3$};
				\node[scale=0.8] at (17.5,0.45){$p$};
				\node[scale=0.78] at (19.25,0.55){$p_3^{-\sigma}$};
				\node[scale=0.8] at (14.5,1){$\infty$};
				\node[scale=0.8] at (22,1){$0$};
			\end{tikzpicture}
		\end{center}
		\caption{The CFT blocks ${\cal F}_{{\text f}}(z,x|k_0^{\sigma },p)$ and ${\cal F}_{{\text b}}(z,x|p,p_3^{-\sigma })$}
		\label{fig:brandfusblocks}
	\end{figure}
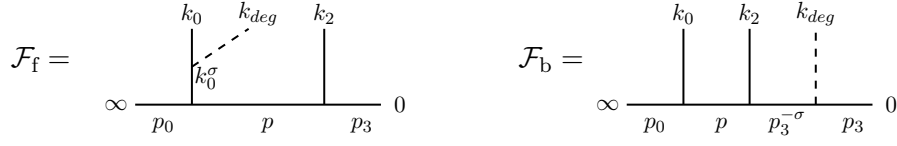
	Obviously, the solutions (\ref{Ganz3}) and (\ref{Ganz4}) are well suited to 
	investigate the regions $z\sim 1$ and $z\sim 0$ respectively. Then the functions 
	\bea 
	{\cal F}_{{\text f}}(z,x)&=&A(z,x)G_{{\text f}}(z,x|k_0^{\sigma},p)
	\\
	{\cal F}_{{\text b}}(z,x)&=&A(z,x)G_{{\text b}}(z,x|p,p_3^{-\sigma})
	\eea
	with $A(z,x)$ given in (\ref{Azx}), are the canonically normalized conformal blocks in the fusion and 
	braiding channels (see (\ref{GvsF}) and  Fig.\ref{fig:brandfusblocks}).
	
\begin{figure}[t]
		\begin{center}
			\begin{tikzpicture}[scale=0.50]
				\draw (1,1)	-- (3,1);
				\draw (3,1)	-- (5,1);
				\draw (5,1)	-- (7,1);
				\draw (7,1)	-- (9,1);
				\draw (3,1) -- (3,3);
				\draw[dashed] (5,1) -- (5,3);
				\draw (7,1) -- (7,3);
				\node[scale=0.8] at (0.5,1){$\infty$};
				\node[scale=0.8] at (9.4,1){$0$};
				\node[scale=0.8] at (3,3.4){$k_0$};
				\node[scale=0.8] at (5,3.4){$k_{deg}$};
				\node[scale=0.8] at (7,3.4){$k_2$};
				\node[scale=0.8] at (2,0.45){$p_0$};
				\node[scale=0.79] at (4,0.55){$p^{-\sigma }$};
				\node[scale=0.8] at (6,0.45){$p$};
				\node[scale=0.8] at (8,0.45){$p_3$};
				\node[scale=0.8] at (2.6,2){$1$};
				\node[scale=0.8] at (4.6,2){$z$};
				\node[scale=0.8] at (6.6,2){$x$};
				\node[scale=0.8] at (10.5,2.2){$=$};
				\node[scale=0.8] at (14,2.2){$ \sum_{\sigma' }	F_{\sigma,\sigma '}\left[\begin{array}{cc}
						k_0& k_{deg}\\
						p_0& p \end{array}\right]$};
				\draw (17,1)	-- (23.5,1);
				\draw (22,1) -- (22,3);
				\draw (18.5,1) -- (18.5,3);
				\draw [dashed] (18.5,2)--(20,3);
				\node[scale=0.8] at (18.5,3.4){$k_0$};
				\node[scale=0.8] at (20.3,3.4){$k_{deg}$};
				\node[scale=0.8] at (22,3.4){$k_2$};
				\node[scale=0.8] at (17.75,0.45){$p_0$};
				\node[scale=0.8] at (20.5,0.45){$p$};
				\node[scale=0.8] at (23,0.45){$p_3$};
				\node[scale=0.8] at (19,1.7){$k_0^{\sigma '}$};
			\end{tikzpicture}
		\end{center}

		\begin{center}
			\begin{tikzpicture}[scale=0.50]
				\draw (1,1)	-- (3,1);
				\draw (3,1)	-- (5,1);
				\draw (5,1)	-- (7,1);
				\draw (7,1)	-- (9,1);
				\draw (3,1) -- (3,3);
				\draw[dashed] (5,1) -- (5,3);
				\draw (7,1) -- (7,3);
				\node[scale=0.8] at (0.5,1){$\infty$};
				\node[scale=0.8] at (9.4,1){$0$};
				\node[scale=0.8] at (3,3.4){$k_0$};
				\node[scale=0.8] at (5,3.4){$k_{deg}$};
				\node[scale=0.8] at (7,3.4){$k_2$};
				\node[scale=0.8] at (2,0.45){$p_0$};
				\node[scale=0.8] at (4,0.45){$p$};
				\node[scale=0.8] at (6,0.55){$p^{\sigma }$};
				\node[scale=0.8] at (8,0.45){$p_3$};
				\node[scale=0.8] at (2.6,2){$1$};
				\node[scale=0.8] at (4.6,2){$z$};
				\node[scale=0.8] at (6.6,2){$x$};
				\node[scale=0.8] at (10.5,2.2){$=$};
				\node[scale=0.8] at (14,2.2){$ \sum_{\sigma' }	B_{\sigma,\sigma '}\left[\begin{array}{cc}
						k_{deg} & k_2\\
						p& p_3 \end{array}\right]$};
				\draw (17,1)	-- (23.5,1);
				\draw (18.5,1) -- (18.5,3);
				\draw (20.25,1) -- (20.25,3);
				\draw [dashed] (22,1) -- (22,3);
				\node[scale=0.8] at (18.5,3.4){$k_0$};
				\node[scale=0.8] at (20.25,3.4){$k_2$};
				\node[scale=0.8] at (22,3.4){$k_{deg}$};
				\node[scale=0.8] at (17.75,0.45){$p_0$};
				\node[scale=0.8] at (23,0.45){$p_3$};
				\node[scale=0.8] at (19.5,0.45){$p$};
				\node[scale=0.78] at (21.25,0.55){$p_3^{-\sigma'}$};
			\end{tikzpicture}
		\end{center}
		\caption{Fusion and braiding operations on conformal block}
		\label{fig:Fusion_Braiding}
	\end{figure}
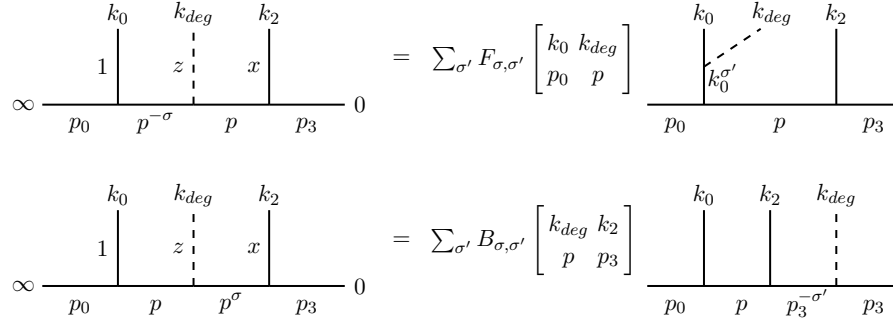

	Using identities (\ref{rule2F11}) and (\ref{rule2F1inf})  one can verify that
	\bea 
	H_{\sigma}^1(z)&=&\sum_{\sigma'=\pm }F_{\sigma,\sigma '}
	\left[\begin{array}{cc}
		k_0& k_{deg}\\
		p_0& p \end{array}\right]{\tilde H}^1_{\sigma '}(z)
	\\
	H_{\sigma}^2(y)&=&\sum_{\sigma'=\pm }	B_{\sigma,\sigma '}\left[\begin{array}{cc}
		k_{deg} & k_2\\
		p& p_3 \end{array}\right]{\tilde H}^2_{\sigma '}(y)
	\eea
	where
	\bea
	\label{Fus} 
	&
	F_{\sigma,\sigma '}
	\left[\begin{array}{cc}
		k_0& k_{deg}\\
		p_0& p \end{array}\right]
	=\frac{\Gamma \left(\sigma ' \left(1-\kappa _2-\kappa _3\right)\right) 
		\Gamma (2 b \sigma p+1)}{\Gamma \left(b \sigma p+\frac{1}{2} 
		(\sigma '+1) \left(1-\kappa _2-\kappa _3\right)+\kappa _3\right) 
		\Gamma \left(b \sigma p+\kappa _2+\frac{1}{2} (\sigma '+1) 
		\left(1-\kappa _2-\kappa _3\right)\right)}
	\qquad \quad
	\\
	\label{Br}
	&
	B_{\sigma,\sigma '}\left[\begin{array}{cc}
		k_{deg} & k_2\\
		p& p_3 \end{array}\right]
	=\frac{e^{i \pi  \left(\frac{b^2}{2}+b \sigma p+\frac{1}{2} 
			\left(\gamma _0-1\right) (\sigma '+1)-\kappa _1\right)}
		\Gamma \left(\sigma ' \left(\gamma _0-1\right)\right) 
		\Gamma (1-2 b \sigma p)}{\Gamma \left(\frac{b^2}{2}-\sigma p b+\frac{1}{2} (\sigma '+1) 
		\left(\gamma _0-1\right)-\kappa _1+1\right) 
		\Gamma \left(-\frac{b^2}{2}-\sigma p b+\frac{1}{2} (\sigma '-1) \left(\gamma _0-1\right)+\kappa _1\right)}\qquad
	\eea
	It is of crucial importance that $P$, ${\hat P}$ are even functions on intermediate 
	momentum $p$. This is why they do not depend on the choice of $\sigma$, which specifies 
	an intermediate channel. Hence, above relations can be readily promoted to the full conformal blocks:
	\bea 
	\label{F_relation}
	{\cal F}(z,x|p^{-\sigma},p)&=&\sum_{\sigma'=\pm }F_{\sigma,\sigma '}
	\left[\begin{array}{cc}
		k_0& k_{deg}\\
		p_0& p \end{array}\right]{\cal F}_{{\text f}}(z,x|k_0^{\sigma '},p)
	\\
	\label{B_relation}
	{\cal F}(z,x|p,p^{\sigma})&=&\sum_{\sigma'=\pm }	B_{\sigma,\sigma '}\left[\begin{array}{cc}
		k_{deg} & k_2\\
		p& p_3 \end{array}\right]{\cal F}_{{\text b}}(z,x|p,p_3^{-\sigma '})
	\eea
	so that (\ref{Fus}) and (\ref{Br}) are just the fusion and braiding matrices  (see Fig.\ref{fig:Fusion_Braiding}) 
	\cite{moore1988polynomial}, previously calculated by investigation of degenerate 
		four-point function in \cite{Teschner:1997ft}. The advantage of our method is that it is 
		mathematically quite rigorous and does not depend on existence of local OPE. Even more, to some extent  it 
		justifies the local OPE conjecture.
	
	\subsection{Four point conformal block}
	\label{4pointBlock}
	The four-point conformal block can be derived using e.g. equation (\ref{Ganz1}) or (\ref{Ganz2})
	by sending the insertion point $z\to \infty $ or $z\to 0$ respectively\footnote 
	{The alternative options taking the limits $z\to 1 $ or $z\to x$ work as well.}.
	To provide a comprehensive understanding of the derivation process, 
	we will illustrate the first case in full detail, the derivation of the second one is quite similar.
	So let us consider the colliding limit
	\bea
	0 \ll x \ll 1
	\quad
	{\rm and }
	\quad
	z\to \infty
	\eea 
	Then from  (\ref{Ganz1}), (\ref{cij}) and (\ref{rule2F1inf}), depending on the choice of singularity 
	at large $z$, we get combination of two branches
	\bea
	\label{largez1}
	\frac{(-1)^{b p \sigma +\kappa _2} \Gamma \left(b p \sigma -\kappa _2+1\right) \Gamma \left(b p \sigma +\kappa _3\right)}{\Gamma \left(\kappa _3-\kappa _2\right) \Gamma (2 b p \sigma +1)}	z^{\kappa _1+\kappa _2}   x^{p^2-\frac{\kappa _1^2}{b^2}} G(z,x|p^{-\sigma},p)
	=\nn\\
	\sum_{i=0}^\infty \left(h_{i,i}-(\kappa_1+\kappa_2)\hat{h}_{i,i}\right)x^i
	\eea
	and
	\bea
	\label{largez2}
	\frac{ (-1)^{b p \sigma +\kappa _3} \Gamma \left(b p \sigma+\kappa _2\right) \Gamma \left(b p \sigma-\kappa _3+1\right)}{ \Gamma \left(\kappa _2-\kappa _3\right) \Gamma (2 b p \sigma+1)}z^{\kappa _1+\kappa _3}x^{p^2-\frac{\kappa _1^2}{b^2}} G(z,x|p^{-\sigma},p)
	=\nn\\
	\sum_{i=0}^\infty \left(h_{i,i}-(\kappa_1+\kappa_3)\hat{h}_{i,i}\right)x^i
	\eea
	It is easy to check that the exponents of $z$ exactly match with those expected from fusion of 
	$V_{k_{deg}}(z)$ with the bra state $\langle p_0|V_{k_{deg}}(z)$, which produces states either $\langle p_0+b/2|$ or 
	$\langle p_0-b/2|$.
	
	Since $\kappa_2$ differs from $\kappa_3$ only by the sign of $p_0$ see (\ref{kappaToP}) it is sufficient 
	to analyse the case (\ref{largez2}).
	Using the recursion relation (\ref{RecrelTr0}) we get
	\bea
	\label{sec_ap_1_inst}
	(1-x)^{\frac{\left(b^2+\kappa _2+\kappa _3\right) \left(\gamma _0-2 \kappa _1-1\right)}{2 b^2}}\sum_{i=0}^\infty \left(h_{i,i}-(\kappa_1+\kappa_3)\hat{h}_{i,i}\right)x^i=
	\\ \nn
	1+\frac{x \left(b^2+Q^2-4 \left(p_0 \left(b-p_0\right)+k_0^2+p^2\right)\right) \left(4 \left(p_3^2-k_2^2-p^2\right)+Q^2\right)}{8  \left(Q^2-4 p^2\right)}+O\left(x^2\right)
	\eea
	We have checked that this result (and also further corrections up to $O(x^7)$) exactly 
	agrees with $4$-point block
	\bea
	\label{tp_to_p_reg}
	& \langle p_0+\frac{b}{2}| V_{k_0} (1) V_{k_2} (x)  |p_3\rangle
	\eea
	obtained by using AGT and instanton counting (\ref{4_point_block_ins}).
	As already mentioned one can derive $4$-point blocks by approaching to other singularities as well,
	e.g. by using (\ref{Ganz2}) and considering the limit
	\bea
	0 \ll z \ll x \ll 1
	\quad
	{\rm so \,\, that}
	\quad
	\frac{x}{z} \to \infty
	\eea
	
	
	\section{Summary}
	\label{summary}
	The main results of the present paper are  solutions of the BPZ differential equation (\ref{BPZ}) 
	expressed in terms of hypergeometric functions. By separating the free-field contribution (\ref{GvsF}), (\ref{Azx}), 
	we represent the BPZ equation in the equivalent form (\ref{BPZren}). This leads to two distinct pairs of solutions, 
	given in (\ref{Ganz1}), (\ref{Ganz2}), (\ref{Ganz3}), and (\ref{Ganz4}).
	Remarkably, the connection matrices relating these solutions can be evaluated in a straightforward way
    using standard hypergeometric identities (see (\ref{F_relation}), (\ref{B_relation})).
	 
	Still, to fully justify our approach with mathematical rigour one should prove that the extra conditions 
	(\ref{extra_conditions}) are indeed satisfied. Though we have checked this up to rather high orders, 
	currently it remains a conjecture.
	
	An immediate application of our results is the construction of physical five-point correlation functions in quantum Liouville theory, obtained by incorporating the DOZZ structure constants  \cite{Dorn:1994xn,Zamolodchikov:1995aa}. It would also be of interest to extend our approach to the ${\cal N}=1$ supersymmetric Liouville case.
	
	Recently irregular blocks attract considerable attention 
	due to their relevance in Argyres Douglas theories \cite{,Argyres:1995jj,Argyres:1995xn} in $\Omega$-background 
	\cite{Gaiotto:2009ma,Gaiotto:2012sf,Nishinaka:2012kn,
		Bonelli:2016qwg,Nishinaka:2019nuy,
		Poghosyan:2023zvy,Poghossian:2025nef}. Let us also mention that in self-dual case 
	$\epsilon_1=-\epsilon_2$ the Nekrasov partition functions are related to $\tau$-functions 
	of  Painlev\'{e} transcendents \cite{Gamayun:2013auu}, 
	\cite{Bonelli:2016qwg,Poghosyan:2023zvy,Bonelli:2025owb}. Note also that 
	the generic 	$\epsilon_1$, $\epsilon_2$ case corresponds to the recently introduced quantum 
	Painlev\'{e} transcendents \cite{Bonelli:2025owb}.

	The technique developed in this paper can be extended to the irregular conformal blocks with 
	degenerate field insertion.  We are going to present our 
	results in this direction in a forthcoming publication.
	
	Another interesting particular case of $\Omega$-background $\epsilon_2=0$, called 
	Nekrasov-Shatashvili limit \cite{Nekrasov:2009rc} (see also \cite{Poghossian:2010pn,Fucito:2011pn}), appears to have unexpected applications in analyzing 
	gravitational perturbations \cite{Aminov:2020yma,Bianchi:2021xpr,Bianchi:2021mft,Fioravanti:2021dce,
		Consoli:2022eey,Fucito:2023afe,Aminov:2023jve,Fucito:2024wlg,DiRusso:2024hmd,Bautista:2023sdf,Liu:2024eut}.    
	In 2d CFT side NS limit is just the quasi-classical limit $b\to 0$, when 
	the BPZ differential equation (\ref{BPZ}) reduces to the Heun equation. This particular limit
	(and also its confluent case) has been analysed in recent paper \cite{Cipriani:2025ikx}, 
	where the results were used to compute gravitational waveform emitted by 
	a particle moving in a Schwarzschild geometry.
	
	\acknowledgments
	
	The research of R.P. was supported by the Armenian SCS grants 21AG-1C060 and 24WS-
	1C031. Similarly, H.P.’s work received support from Armenian SCS grants 21AG-1C062 and
	24WS-1C031.
	
	\begin{appendix}
		\section{Instanton counting and AGT}
		\label{instAGT}
		Let us consider ${\cal N}=2$ supersymmetric $U(2)^n$ linear quiver gauge theory 
		in $\Omega$-background. The initial $i=0$ and the last $i=n+1$ nodes of the quiver correspond to fundamental 
		and anti-fundamental hypermultiplets charged with respect to neighboring $i=1$ and $i=n$ gauge nodes. The 
		intermediate links represent bi-fundamental hypermultiplets. It is well known that the instanton part of the 
		partition function can be represented as
		\begin{small}
			\begin{eqnarray}
				\label{instpartition}
				Z_{inst}(\vec{a}_0,\vec{a}_1, \dots ,\vec{a}_{n+1}|q_1,\dots,q_n)=
				\sum_{\{\vec{Y}_1,\dots,\vec{Y}_{n+1}\}}F_{\vec{Y}_{1},\dots,\vec{Y}_n}\left(\vec{a}_{0}, \dots , \vec{a}_{n+1}\right)
				q_1^{|\vec{Y}_1|}q_2^{|\vec{Y}_2|} \dots q_n^{|\vec{Y}_n|}
			\end{eqnarray}
		\end{small}
		where the sum goes over all possible $n$-tuples of Young diagrams $\vec{Y}_i=\{Y_{i,1},Y_{i,2}\}$,
		$i=1,2,\dots,n$. It is convenient also to associate with nodes $i=0$ and $i=n+1$ 
		pairs of empty diagrams $\{\varnothing,\varnothing\}$. 
		$|\vec{Y}_i|$ denotes the total number of boxes in pair of diagrams $Y_{i,1}$, $Y_{i,2}$,
		$q_i$'s are the instanton counting parameters and $\vec{a}_i=(a_{i,1},a_{i,2})$ encode the expectation values of scalar fields in vector multiplets and the hypermultiplet masses. Namely $a_{i,1}-a_{i,2}\over 2$, $i=1,2,\ldots,n$ are the expectation values of $i$-th scalar and 
		$a_{i,1}+a_{i,2}-a_{i-1,1}-a_{i-1,2}\over 2$, $i=0,1,\ldots,n+1$ are the hypermultiplet masses. Simultaneous shift of all $\vec{a}_i$'s by a constant amount does not affect physics, hence, one is free to choose e.g. $a_{2,1}=-a_{2,2}$. The expansion coefficients in (\ref{instpartition})
		have a factorized form
		\begin{eqnarray}
			\label{F}
			& F_{\vec{Y}_1,\dots,\vec{Y}_n}=
			\displaystyle \prod_{i,j=1}^2 
			\prod_{l=0}^n 
			\frac{Z_{bf}(a_{l,i}, Y_{l,i} \mid  a_{l+1,j},Y_{l+1,j})  }
			{Z_{bf}(a_{l,i},Y_{l,i}\mid   a_{l,j},Y_{l,j})  }
		\end{eqnarray}
		where
		\begin{eqnarray}
			\label{Zbf}
			&Z_{bf}(a,\lambda\mid b,\mu)=\\
			&\displaystyle\prod_{s\in\lambda}\big(a-b-\epsilon_1L_{\mu}(s)+
			\epsilon_2(1+A_{\lambda}(s))\big)
			\displaystyle\prod_{s\in\mu}\big(a-b+\epsilon_1(1+L_{\lambda}(s))
			-\epsilon_2A_{\mu}(s)\big)\nonumber
		\end{eqnarray}
		As usual $\epsilon_{1,2}$ denote the $\Omega$-background parameters and 
		$A_{\lambda}(s)$, $L_{\lambda}(s)$ are the arm and leg lengths of the box $s$ 
		with respect to the diagram $\lambda$.
		Notice that $Z_{bf}(a_{0,i},Y_{0,i}\mid   a_{0,j},Y_{0,j})=1 $.
		\subsection{Five point conformal block and AGT}
		In this paper we deal with quiver gauge theory with number of nodes $n=1$ or $n=2$. 
		Let us first  review the AGT correspondence \cite{Alday:2009aq} for the case of two nodes.
		\begin{figure}
			\begin{center}
				\begin{tikzpicture}[scale=0.50]
					\draw (1,1)	-- (3,1);
					\draw (3,1)	-- (5,1);
					\draw (5,1)	-- (7,1);
					\draw (7,1)	-- (9,1);
					\draw (3,1) -- (3,3);
					\draw[dashed] (5,1) -- (5,3);
					\draw (7,1) -- (7,3);
					\node[scale=0.8] at (0.5,1){$\infty$};
					\node[scale=0.8] at (9.4,1){$0$};
					\node[scale=0.8] at (3,3.4){$k_0$};
					\node[scale=0.8] at (5,3.4){$k_{deg}$};
					\node[scale=0.8] at (7,3.4){$k_2$};
					\node[scale=0.8] at (2,0.6){$p_0$};
					\node[scale=0.8] at (4,0.6){$p_1$};
					\node[scale=0.8] at (6,0.6){$p_2$};
					\node[scale=0.8] at (8,0.6){$p_3$};
					\node[scale=0.8] at (2.6,2){$1$};
					\node[scale=0.8] at (4.6,2){$z$};
					\node[scale=0.8] at (6.6,2){$x$};
					\draw [draw=black] (12.5,1.5) rectangle (14,3);
					\draw (14,2.25)	-- (15,2.25);
					\draw[color=black] (15.75,2.25) circle (0.75);
					\draw (16.5,2.25)	-- (17.5,2.25);
					\draw[color=black] (18.25,2.25) circle (0.75);
					\draw (19,2.25)	-- (20,2.25);
					\draw [draw=black] (20,1.5) rectangle (21.5,3);
					\node[scale=0.65] at (15.75,2.25){$U(2)$};
					\node[scale=0.65] at (18.25,2.25){$U(2)$};
					\node[scale=0.85] at (15.75,3.5){$q_1$};
					\node[scale=0.85] at (18.25,3.5){$q_2$};
					\node[scale=0.8] at (13.25,1){$\vec{a}_0$};
					\node[scale=0.8] at (15.75,1){$\vec{a}_1$};
					\node[scale=0.8] at (18.25,1){$\vec{a}_2$};
					\node[scale=0.8] at (20.75,1){$\vec{a}_3$};
					\node[scale=0.8] at (11,2.2){$\Longleftrightarrow$};
				\end{tikzpicture}
			\end{center}
			\caption{AGT duality for $5$-point conformal block;
				relations between the parameters are given in 
				(\ref{vertmom}), (\ref{hormom}), (\ref{agt_ins_point}).}
			\label{fig:5pblockAGT}
		\end{figure}
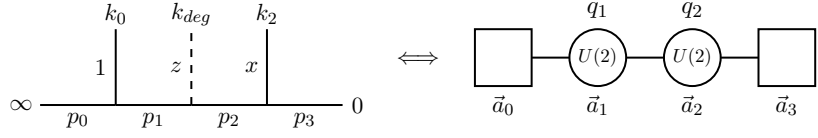
		The relation between the instanton partition function of gauge theory and the 
		2d Liouville  five-point conformal  block (\ref{5pointCB}) reads
		\bea
		\label{AGT_F_Z_inst}
		{\cal F}(z,x)=x^{h_{p_2}-h_{k_2}-h_{p_3}}z^{h_{p_1}-h_{k_{deg}}-h_{p_2}}\frac{Z_{inst}}{Z^{U(1)}}
		\eea
		The $U(1)$ factor is given by
		\bea
		\label{U1factor}
		& Z^{U(1)}=(1-z)^{\left(\frac{Q}{2}-k_0\right)\left(Q+2k_{deg}\right)}(1-x)^{\left(\frac{Q}{2}-k_0\right)\left(Q+2k_{2}\right)}
		\left(1-\frac{x}{z}\right)^{\left(\frac{Q}{2}-k_{deg}\right)\left(Q+2k_{2}\right)}
		\eea
		The map between  gauge parameters in (\ref{instpartition}) and CFT parameters in (\ref{AGT_F_Z_inst})
		can  be established using the rules:
		\begin{itemize}
			\item
			The differences between the “centers of masses” ${\bar a}_i=(a_{i,1}+a_{i,2})/2$ of the successive  nodes
			give the charges of the ``vertical"  entries of the conformal block 
			\bea
			\label{vertmom}
			& \bar{a}_{1}-\bar{a}_{0}=\frac{Q}{2}- k_0, \quad \bar{a}_{2}-\bar{a}_{1}=\frac{Q}{2}- k_{deg}, \, \quad
			\, \bar{a}_{3}-\bar{a}_{2}=\frac{Q}{2}-k_2
			\eea
			\item
			The gauge parameters with the subtracted centers of masses give the momenta of the ``horizontal" entries of the conformal block:
			\begin{small}
				\bea
				\label{hormom}
				&\frac{a_{0,1}-a_{0,2}}{2}= p_0\,,
				\quad
				\frac{a_{1,1}-a_{1,2}}{2}= p_1\, ,
				\quad
				\frac{a_{2,1}-a_{2,2}}{2}= p_2\,,
				\quad
				\frac{a_{3,1}-a_{3,2}}{2}= p_3
				\eea
			\end{small}
		\end{itemize}
		We also identify 
		\bea
		\label{agt_ins_point}
		& q_1=z\,,\quad
		q_2=\frac{x}{z}\,,\quad
		\epsilon _1=b\,,\quad
		\epsilon _2=\frac{1}{b}
		\eea
		\subsection{Four point conformal block and AGT}
		\label{rank0AGT}
		Here we compute the four-point conformal block using instanton counting and AGT relation
		in order to compare it with the result obtained through the recursive approach described in this paper.
		
		The AGT map for the one node case is
		\begin{small}
			\bea
			a_{0,1}&=&k_0+p_0-\frac{Q}{2}\,,
			\qquad\quad\,\,
			a_{0,2}=k_0-p_0-\frac{Q}{2}\,,
			\quad
			a_{1,2}=-a_{1,1}=-p\,,\nn
			\\
			a_{2,1}&=&\frac{1}{2} \left(Q-2 k_2+2 p_3\right)\,,
			\quad
			a_{2,2}=\frac{1}{2} \left(Q-2 k_2-2 p_3\right)
			\eea
		\end{small}
		and
		\bea
		Z^{U(1)}_{1-node}=(1-q_1)^{\frac{1}{\epsilon _1 \epsilon _2}(a_{1,1}+a_{1,2}-a_{0,1}-a_{0,2})(\epsilon _1+\epsilon _2-\frac{1}{2} \left(a_{2,1}+a_{2,2}-a_{1,1}-a_{1,2}\right))}
		\eea
		These relations together with (\ref{instpartition}) specified for one node gives
		\bea
		\label{4_point_block_ins}
		&\frac{Z_{inst}}{Z^{U(1)}_{1-node}}=
		1+\frac{x \left(Q^2-4  k_0^2-4  p^2+4  p_0^2\right) \left(Q^2-4  k_2^2-4  p^2+4  p_3^2\right)}{8  \left(Q^2-4  p^2\right) }+O(x^2)\,,
		\eea
		which coincides with (\ref{sec_ap_1_inst}).
	\end{appendix}

\bibliographystyle{JHEP}

\begin{thebibliography}{10}
	
	\bibitem{Seiberg:1994rs}
	N.~Seiberg and E.~Witten, {\it {Electric - magnetic duality, monopole
			condensation, and confinement in N=2 supersymmetric Yang-Mills theory}},
	{\em Nucl. Phys. B} {\bf 426} (1994) 19--52,
	[\href{http://arxiv.org/abs/hep-th/9407087}{{\tt hep-th/9407087}}]. [Erratum:
	Nucl.Phys.B 430, 485--486 (1994)].
	
	\bibitem{Seiberg:1994aj}
	N.~Seiberg and E.~Witten, {\it {Monopoles, duality and chiral symmetry breaking
			in N=2 supersymmetric QCD}},  {\em Nucl. Phys. B} {\bf 431} (1994) 484--550,
	[\href{http://arxiv.org/abs/hep-th/9408099}{{\tt hep-th/9408099}}].
	
	\bibitem{Nekrasov:2002qd}
	N.~A. Nekrasov, {\it {Seiberg-Witten prepotential from instanton counting}},
	{\em Adv. Theor. Math. Phys.} {\bf 7} (2003), no.~5 831--864,
	[\href{http://arxiv.org/abs/hep-th/0206161}{{\tt hep-th/0206161}}].
	
	\bibitem{Flume:2002az}
	R.~Flume and R.~Poghossian, {\it {An Algorithm for the microscopic evaluation
			of the coefficients of the Seiberg-Witten prepotential}},  {\em Int. J. Mod.
		Phys. A} {\bf 18} (2003) 2541,
	[\href{http://arxiv.org/abs/hep-th/0208176}{{\tt hep-th/0208176}}].
	
	\bibitem{Bruzzo:2002xf}
	U.~Bruzzo, F.~Fucito, J.~F. Morales, and A.~Tanzini, {\it {Multiinstanton
			calculus and equivariant cohomology}},  {\em JHEP} {\bf 05} (2003) 054,
	[\href{http://arxiv.org/abs/hep-th/0211108}{{\tt hep-th/0211108}}].
	
	\bibitem{Nekrasov:2003rj}
	N.~Nekrasov and A.~Okounkov, {\it {Seiberg-Witten theory and random
			partitions}},  {\em Prog. Math.} {\bf 244} (2006) 525--596,
	[\href{http://arxiv.org/abs/hep-th/0306238}{{\tt hep-th/0306238}}].
	
	\bibitem{Flume:2004rp}
	R.~Flume, F.~Fucito, J.~F. Morales, and R.~Poghossian, {\it {Matone's relation
			in the presence of gravitational couplings}},  {\em JHEP} {\bf 04} (2004)
	008, [\href{http://arxiv.org/abs/hep-th/0403057}{{\tt hep-th/0403057}}].
	
	\bibitem{Gaiotto:2009we}
	D.~Gaiotto, {\it {N=2 dualities}},  {\em JHEP} {\bf 08} (2012) 034,
	[\href{http://arxiv.org/abs/0904.2715}{{\tt arXiv:0904.2715}}].
	
	\bibitem{Alday:2009aq}
	L.~F. Alday, D.~Gaiotto, and Y.~Tachikawa, {\it {Liouville Correlation
			Functions from Four-dimensional Gauge Theories}},  {\em Lett. Math. Phys.}
	{\bf 91} (2010) 167--197, [\href{http://arxiv.org/abs/0906.3219}{{\tt
			arXiv:0906.3219}}].
	
	\bibitem{Poghossian:2009mk}
	R.~Poghossian, {\it {Recursion relations in CFT and N=2 SYM theory}},  {\em
		JHEP} {\bf 12} (2009) 038, [\href{http://arxiv.org/abs/0909.3412}{{\tt
			arXiv:0909.3412}}].
	
	\bibitem{Gaiotto:2012sf}
	D.~Gaiotto and J.~Teschner, {\it {Irregular singularities in Liouville theory
			and Argyres-Douglas type gauge theories, I}},  {\em JHEP} {\bf 12} (2012)
	050, [\href{http://arxiv.org/abs/1203.1052}{{\tt arXiv:1203.1052}}].
	
	\bibitem{Belavin:1984vu}
	A.~A. Belavin, A.~M. Polyakov, and A.~B. Zamolodchikov, {\it {Infinite
			Conformal Symmetry in Two-Dimensional Quantum Field Theory}},  {\em Nucl.
		Phys. B} {\bf 241} (1984) 333--380.
	
	\bibitem{moore1988polynomial}
	G.~Moore and N.~Seiberg, {\it Polynomial equations for rational conformal field
		theories},  {\em Physics Letters B} {\bf 212} (1988), no.~4 451--460.
	
	\bibitem{Teschner:1997ft}
	J.~Teschner, {\it {On structure constants and fusion rules in the SL(2,C) /
			SU(2) WZNW model}},  {\em Nucl. Phys. B} {\bf 546} (1999) 390--422,
	[\href{http://arxiv.org/abs/hep-th/9712256}{{\tt hep-th/9712256}}].
	
	\bibitem{Dorn:1994xn}
	H.~Dorn and H.~J. Otto, {\it {Two and three point functions in Liouville
			theory}},  {\em Nucl. Phys. B} {\bf 429} (1994) 375--388,
	[\href{http://arxiv.org/abs/hep-th/9403141}{{\tt hep-th/9403141}}].
	
	\bibitem{Zamolodchikov:1995aa}
	A.~B. Zamolodchikov and A.~B. Zamolodchikov, {\it {Structure constants and
			conformal bootstrap in Liouville field theory}},  {\em Nucl. Phys. B} {\bf
		477} (1996) 577--605, [\href{http://arxiv.org/abs/hep-th/9506136}{{\tt
			hep-th/9506136}}].
	
	\bibitem{Argyres:1995jj}
	P.~C. Argyres and M.~R. Douglas, {\it {New phenomena in SU(3) supersymmetric
			gauge theory}},  {\em Nucl. Phys. B} {\bf 448} (1995) 93--126,
	[\href{http://arxiv.org/abs/hep-th/9505062}{{\tt hep-th/9505062}}].
	
	\bibitem{Argyres:1995xn}
	P.~C. Argyres, M.~R. Plesser, N.~Seiberg, and E.~Witten, {\it {New N=2
			superconformal field theories in four-dimensions}},  {\em Nucl. Phys. B} {\bf
		461} (1996) 71--84, [\href{http://arxiv.org/abs/hep-th/9511154}{{\tt
			hep-th/9511154}}].
	
	\bibitem{Gaiotto:2009ma}
	D.~Gaiotto, {\it {Asymptotically free $\mathcal{N} = 2$ theories and irregular
			conformal blocks}},  {\em J. Phys. Conf. Ser.} {\bf 462} (2013), no.~1
	012014, [\href{http://arxiv.org/abs/0908.0307}{{\tt arXiv:0908.0307}}].
	
	\bibitem{Nishinaka:2012kn}
	T.~Nishinaka and C.~Rim, {\it {Matrix models for irregular conformal blocks and
			Argyres-Douglas theories}},  {\em JHEP} {\bf 10} (2012) 138,
	[\href{http://arxiv.org/abs/1207.4480}{{\tt arXiv:1207.4480}}].
	
	\bibitem{Bonelli:2016qwg}
	G.~Bonelli, O.~Lisovyy, K.~Maruyoshi, A.~Sciarappa, and A.~Tanzini, {\it {On
			Painlev\'e/gauge theory correspondence}},  {\em Lett. Matth. Phys.} {\bf 107}
	(2017), no.~12 2359--2413, [\href{http://arxiv.org/abs/1612.06235}{{\tt
			arXiv:1612.06235}}].
	
	\bibitem{Nishinaka:2019nuy}
	T.~Nishinaka and T.~Uetoko, {\it {Argyres-Douglas theories and Liouville
			Irregular States}},  {\em JHEP} {\bf 09} (2019) 104,
	[\href{http://arxiv.org/abs/1905.03795}{{\tt arXiv:1905.03795}}].
	
	\bibitem{Poghosyan:2023zvy}
	H.~Poghosyan and R.~Poghossian, {\it {A note on rank 5/2 Liouville irregular
			block, Painlev\'e I and the $ \mathcal{H} _{0}$ Argyres-Douglas theory}},
	{\em JHEP} {\bf 11} (2023) 198, [\href{http://arxiv.org/abs/2308.09623}{{\tt
			arXiv:2308.09623}}].
	
	\bibitem{Poghossian:2025nef}
	R.~Poghossian and H.~Poghosyan, {\it {A note on rank $\frac{3}{2}$ Liouville
			irregular block}},  \href{http://arxiv.org/abs/2502.10169}{{\tt
			arXiv:2502.10169}}.
	
	\bibitem{Gamayun:2013auu}
	O.~Gamayun, N.~Iorgov, and O.~Lisovyy, {\it {How instanton combinatorics solves
			Painlev\'e VI, V and IIIs}},  {\em J. Phys. A} {\bf 46} (2013) 335203,
	[\href{http://arxiv.org/abs/1302.1832}{{\tt arXiv:1302.1832}}].
	
	\bibitem{Bonelli:2025owb}
	G.~Bonelli, A.~Shchechkin, and A.~Tanzini, {\it {Refined Painlev\'e/gauge
			theory correspondence and quantum tau functions}},
	\href{http://arxiv.org/abs/2502.01499}{{\tt arXiv:2502.01499}}.
	
	\bibitem{Nekrasov:2009rc}
	N.~A. Nekrasov and S.~L. Shatashvili, {\it {Quantization of Integrable Systems
			and Four Dimensional Gauge Theories}},  in {\em {16th International Congress
			on Mathematical Physics}}, pp.~265--289, 8, 2009.
	\newblock \href{http://arxiv.org/abs/0908.4052}{{\tt arXiv:0908.4052}}.
	
	\bibitem{Poghossian:2010pn}
	R.~Poghossian, {\it {Deforming SW curve}},  {\em JHEP} {\bf 04} (2011) 033,
	[\href{http://arxiv.org/abs/1006.4822}{{\tt arXiv:1006.4822}}].
	
	\bibitem{Fucito:2011pn}
	F.~Fucito, J.~F. Morales, D.~R. Pacifici, and R.~Poghossian, {\it {Gauge
			theories on $\Omega$-backgrounds from non commutative Seiberg-Witten
			curves}},  {\em JHEP} {\bf 05} (2011) 098,
	[\href{http://arxiv.org/abs/1103.4495}{{\tt arXiv:1103.4495}}].
	
	\bibitem{Aminov:2020yma}
	G.~Aminov, A.~Grassi, and Y.~Hatsuda, {\it {Black Hole Quasinormal Modes and
			Seiberg\textendash{}Witten Theory}},  {\em Annales Henri Poincare} {\bf 23}
	(2022), no.~6 1951--1977, [\href{http://arxiv.org/abs/2006.06111}{{\tt
			arXiv:2006.06111}}].
	
	\bibitem{Bianchi:2021xpr}
	M.~Bianchi, D.~Consoli, A.~Grillo, and J.~F. Morales, {\it {QNMs of branes, BHs
			and fuzzballs from quantum SW geometries}},  {\em Phys. Lett. B} {\bf 824}
	(2022) 136837, [\href{http://arxiv.org/abs/2105.04245}{{\tt
			arXiv:2105.04245}}].
	
	\bibitem{Bianchi:2021mft}
	M.~Bianchi, D.~Consoli, A.~Grillo, and J.~F. Morales, {\it {More on the SW-QNM
			correspondence}},  {\em JHEP} {\bf 01} (2022) 024,
	[\href{http://arxiv.org/abs/2109.09804}{{\tt arXiv:2109.09804}}].
	
	\bibitem{Fioravanti:2021dce}
	D.~Fioravanti and D.~Gregori, {\it {A new method for exact results on
			Quasinormal Modes of Black Holes}},
	\href{http://arxiv.org/abs/2112.11434}{{\tt arXiv:2112.11434}}.
	
	\bibitem{Consoli:2022eey}
	D.~Consoli, F.~Fucito, J.~F. Morales, and R.~Poghossian, {\it {CFT description
			of BH\textquoteright{}s and ECO\textquoteright{}s: QNMs, superradiance,
			echoes and tidal responses}},  {\em JHEP} {\bf 12} (2022) 115,
	[\href{http://arxiv.org/abs/2206.09437}{{\tt arXiv:2206.09437}}].
	
	\bibitem{Fucito:2023afe}
	F.~Fucito and J.~F. Morales, {\it {Post Newtonian emission of gravitational
			waves from binary systems: a gauge theory perspective}},  {\em JHEP} {\bf 03}
	(2024) 106, [\href{http://arxiv.org/abs/2311.14637}{{\tt arXiv:2311.14637}}].
	
	\bibitem{Aminov:2023jve}
	G.~Aminov, P.~Arnaudo, G.~Bonelli, A.~Grassi, and A.~Tanzini, {\it {Black hole
			perturbation theory and multiple polylogarithms}},  {\em JHEP} {\bf 11}
	(2023) 059, [\href{http://arxiv.org/abs/2307.10141}{{\tt arXiv:2307.10141}}].
	
	\bibitem{Fucito:2024wlg}
	F.~Fucito, J.~F. Morales, and R.~Russo, {\it {Gravitational wave forms for
			extreme mass ratio collisions from supersymmetric gauge theories}},  {\em
		Phys. Rev. D} {\bf 111} (2025), no.~4 044054,
	[\href{http://arxiv.org/abs/2408.07329}{{\tt arXiv:2408.07329}}].
	
	\bibitem{DiRusso:2024hmd}
	G.~Di~Russo, F.~Fucito, and J.~F. Morales, {\it {Tidal resonances for
			fuzzballs}},  {\em JHEP} {\bf 04} (2024) 149,
	[\href{http://arxiv.org/abs/2402.06621}{{\tt arXiv:2402.06621}}].
	
	\bibitem{Bautista:2023sdf}
	Y.~F. Bautista, G.~Bonelli, C.~Iossa, A.~Tanzini, and Z.~Zhou, {\it {Black hole
			perturbation theory meets CFT2: Kerr-Compton amplitudes from
			Nekrasov-Shatashvili functions}},  {\em Phys. Rev. D} {\bf 109} (2024), no.~8
	084071, [\href{http://arxiv.org/abs/2312.05965}{{\tt arXiv:2312.05965}}].
	
	\bibitem{Liu:2024eut}
	P.~Liu and R.-D. Zhu, {\it {Notes on Quasinormal Modes of charged de Sitter
			Blackholes from Quiver Gauge Theories}},
	\href{http://arxiv.org/abs/2412.18359}{{\tt arXiv:2412.18359}}.
	
	\bibitem{Cipriani:2025ikx}
	A.~Cipriani, G.~Di~Russo, F.~Fucito, J.~F. Morales, H.~Poghosyan, and
	R.~Poghossian, {\it {Resumming Post-Minkowskian and Post-Newtonian
			gravitational waveform expansions}},
	\href{http://arxiv.org/abs/2501.19257}{{\tt arXiv:2501.19257}}.
	
\end{thebibliography}
\providecommand{\href}[2]{#2}\begingroup\raggedright\endgroup
\end{document}